\documentclass[a4paper,fleqn,usenatbib,useAMS]{mnras}

\usepackage{graphicx}	
\usepackage{amsmath}	
\usepackage{amssymb}	
\usepackage{multicol}        
\usepackage{bm}		
\usepackage{pdflscape}	

\usepackage[T1]{fontenc}
\usepackage{ae,aecompl}

\title[electrons' energy in GRB afterglows]{Electrons' energy in GRB afterglows implied by radio peaks}
\author[P. Beniamini \& A. J. van der Horst]{Paz Beniamini, Alexander J. van der Horst
	\\
	Department of Physics, The George Washington University, Washington, DC 20052, USA \\
	Astronomy, Physics and Statistics Institute of Sciences (APSIS)}

\begin{document}
	\label{firstpage}
	\pagerange{\pageref{firstpage}--\pageref{lastpage}}
	\maketitle
\vspace{-0.3cm}	
\begin{abstract}
Gamma-ray burst (GRB) afterglows have been observed across the electromagnetic spectrum, and physical parameters of GRB jets and their surroundings have been derived using broadband modeling. While well-sampled lightcurves across the broadband spectrum are necessary to constrain all the physical parameters, some can be strongly constrained by the right combination of just a few observables, almost independently of the other unknowns. We present a method involving the peaks of radio lightcurves to constrain the fraction of shock energy that resides in electrons, $\epsilon_e$. This parameter is an important ingredient for understanding the microphysics of relativistic shocks; Based on a sample of 36 radio afterglows, we find $\epsilon_e$ has a narrow distribution centered around $0.13-0.15$. Our method is suggested as a diagnostic tool for determining $\epsilon_e$, and to help constrain the broadband modeling of GRB afterglows. Some earlier measurements of the spreads in parameter values for $\epsilon_e$, the kinetic energy of the shock, and the density of the circumburst medium, based on broadband modeling across the entire spectrum, are at odds with our analysis of radio peaks. This could be due to different modeling methods and assumptions, and possibly missing ingredients in past and current modeling efforts. Furthermore, we show that observations at $\gtrsim10$~GHz performed $0.3-30$ days after the GRB trigger, are best suited for pinpointing the synchrotron peak frequency, and consequently $\epsilon_e$. At the same time, observations at lower radio frequencies can pin down the synchrotron self-absorption frequency and help constrain the other physical parameters of GRB afterglows.
\end{abstract}

\vspace{-0.5cm}
\section{Introduction}
\label{sec:introduction}
Since the discovery of multi-wavelength emission from Gamma-Ray Bursts \citep[GRBs;][]{Costa1997,VanParadijs1997,Frail1997}, broadband modeling of their afterglows has proven to be very valuable in studying GRBs and their environments \citep{WijersGalama1999,Chevalier1999,PK2001}. While early multi-wavelength studies were mostly dominated by optical observations, and only a few GRBs had well-sampled light curves in other parts of the electromagnetic spectrum, the {\it Swift} mission was a game changer in this respect \citep{Gehrels2004}. The X-ray light curves provided by {\it Swift}, starting within the first minute after a $\gamma$-ray trigger, sometimes combined with early optical observations from robotic telescopes, enabled a new view on the diversity of GRB afterglow behavior \citep{Nousek2006,OBrien2006}. The {\it Fermi Gamma-ray Space Telescope} added the high-energy gamma-ray component to afterglow studies, getting closer to a full multi-wavelength picture \citep{Abdo2009a,Abdo2009b}. At the other side of the spectrum, most GRB afterglow studies have historically been done with a small number of radio telescopes \citep{Frail1997,Galama1998,Kulkarni1998}, and the number of well-studied radio afterglows is small compared to the optical and X-ray regimes, due to sensitivity constraints \citep{CF2012}. This situation has changed since the upgrade of the Very Large Array \citep{Laskar2013,Zauderer2013}, and the advent of dedicated large GRB programs on facilities such as the Arcminute Microkelvin Imager Large Array \citep[AMI-LA;][]{Stewart2013,Anderson2014}; and this will improve even further with new facilities such as MeerKAT and the Square Kilometer Array in the future \citep{Ghirlanda2013,Burlon2015}.

Even though the radio regime has been hard to probe observationally for GRBs, those observations have been crucial in completing the broadband spectrum and constraining physical parameters of the GRB outflow and the circumburst environment \citep[see][for a recent review]{GvdH2014}. Radio light curves at various observing frequencies can pin down the synchrotron spectral peak and the self-absorption frequency, and their evolution. More interestingly, in some cases radio observations followed the evolution of the GRB outflow until it transitioned from relativistic to non-relativistic expansion \citep{Frail2000,VanDerHorst2008}. Interstellar scintillation at low radio frequencies can cause some difficulty in broadband modeling, but does allow for independent constraints on the source size \citep{Frail1997,Goodman1997}. All these different spectral, dynamic, and modulation effects can make the modeling and interpretation of radio light curves seemingly quite complex. However, given that there are now more than 20 years worth of GRB radio observations, and given the recent telescope upgrades and new radio facilities planned or coming on line in the near future, there are possibilities to explore improvements on the modeling of radio observations. While some new modeling codes allow for more detailed studies of radio observations in the multi-wavelength context \citep{VanEerten2012,VanEerten2015}, in this paper we focus on which physical parameters can be constrained by just radio observables.

Besides the developments in radio astronomy, the work presented in this paper is inspired by earlier studies that focus on deriving strong constraints on certain physical parameters by a few observables in one or two ranges of the electromagnetic spectrum. \citet{Nava2014} showed that the high-energy gamma-ray light curves are clustered when corrected for the gamma-ray prompt emission energy, and showed clustering in derived values for two physical parameters: the fraction of shock energy that resides in electrons, $\epsilon_e$, and the efficiency of the prompt emission mechanism, $\epsilon_{\gamma}$. \cite{Beniamini2015} combined high-energy gamma-ray light curves with X-ray light curves, to put strong constraints on the energetics and the fraction of shock energy that resides in the magnetic field, $\epsilon_B$. In the work presented here we focus on $\epsilon_e$ and how its value can be derived from the peak flux and peak time in a radio light curve. We show that this can be done relatively independently from other physical parameters, making radio light curve peaks very good probes of $\epsilon_e$ and providers of tests for its universality. This is important feedback for numerical simulations of relativistic shocks and their predictions for $\epsilon_e$.

For this study we use a current sample of radio light curves and their peaks, which is described in Section~\ref{sec:Sample}. In Section~\ref{sec:interpretation} we lay out the interpretation of radio light curve peaks, and in Section~\ref{sec:implications} we show the results of our analysis of the radio peak sample. We give an outlook to the future in Section~\ref{sec:future}, and discuss all our results in Section~\ref{sec:discussion}. We summarize and conclude in Section~\ref{sec:conclusions}.

\vspace{-0.3cm}
\section{Radio sample}
\label{sec:Sample}
\cite{CF2012} have built a catalog of radio afterglows of GRBs detected between 1997 and 2011, with most of the available data at 8.5~GHz. The 8.5~GHz lightcurve typically peaks within $\sim 3-6$ days in the source frame ($\sim$10~days in the observer frame), reaching a maximum luminosity of $L_{\nu}\sim 5\times 10^{30} \mbox{ergs} / \mbox{sec} / \mbox{Hz}$. The sample that \cite{CF2012} considered for further detailed analysis included 54 bursts for which radio peaks could be well determined. We consider here a sub-set of that sample, consisting of those bursts with measurements at $\sim$8~GHz, with known redshift, and which are expected to be still relativistic at the time of the peak (where the Lorentz factor satisfies $\Gamma (t_p)\gtrsim 2$). Furthermore, we require that the isotropic equivalent energy released during the prompt emission phase, $E_{\gamma,\rm iso}$, has been well determined. The resulting set consists of 36 GRBs, 34 of which are part of the set considered by \cite{BD2014}, who considered 38 bursts in total. \cite{BD2014} used the catalog of \cite{CF2012} in order to obtain the distributions of $\epsilon_B$. The main difference in the selection criteria is that we require bursts that have a determination of $E_{\gamma,\rm iso}$. As will be detailed in Section~\ref{sec:implications}, this is in fact not essential for our purposes, but minimizes the number of unknowns without significantly affecting the sample size. 
We note that 21 of the 36 bursts ($58\%$) in our sample have measured jet break times \citep{CF2012}, i.e. the time at which the light curve behavior changes because the collimated nature of the outflow becomes noticeable. Two others have upper limits on the jet break time prior to the observing time $t_p$, so the jet break time has occurred before $t_p$ in at least $64\%$ of the GRBs in our sample.
The data used for the sample of GRBs considered in this work is summarized in Table~\ref{tbl:sample}.

\begin{table*}
	\begin{center}
		\caption{Sample of GRBs with radio peaks used in this paper. }
		\begin{tabular}{ccccccccc} \hline \hline
			GRB & $\nu_p$(Hz) & $F_{\nu_p}(\mu Jy)$ & $t_p$(days) & $t_j$ (days) & $z$ & $E_{\gamma,\rm iso,53}$ & $\Psi_{\rm ISM}$  & $\Psi_{\rm wind}$\\ \hline 
			090313 & 8.46 & $435\pm22$ & $9.4\pm 0.5$ & 0.79 & 3.4 & 4.6 & $0.05\pm 0.002$ & $0.05\pm 0.001$ \\
			090423 & 8.46 & $50\pm11$ & $33.1\pm 6.3$ & >45 & 8.19 & 1.1 & $0.14\pm 0.02$ & $0.13\pm 0.02$ \\
			050904 & 8.46 & $76\pm14$ & $35.3\pm 1.5$ & - & 6.3 & 13 & $0.32\pm 0.03$ & $0.13\pm 0.03$ \\
			021004 & 8.46 & $780\pm23$ & $18.7\pm 0.5$ & 7.6 & 2.32 & 0.38 & $0.07\pm 0.001$ & $0.07\pm0.001$ \\
			050603 & 8.46 & $377\pm53$ & $14.1\pm 1.8$& - & 2.82 & 5 & $0.13\pm0.01$ & $0.07\pm0.01$ \\
			000926 & 8.46 & $629\pm24$ & $12.1\pm 0.5$& 1.8 & 2.04 & 2.7 & $0.08\pm0.002$ & $0.07\pm 0.002$ \\
			000301C & 8.46 & $520\pm24$ & $14.1\pm 0.5$ & 5.5 & 2.03 & 0.44 & $0.08\pm0.002$ & $0.08\pm0.002$ \\
			070125 & 8.46 & $1028\pm16$ & $84.1\pm 2$ & 3.8 & 1.55 & 9.6 & $0.35\pm 0.005$ & $0.18\pm 0.004$ \\
			090715B & 8.46 & $191\pm36$ & $9.2\pm 2.3$ & 0.21 & 3 & 2.4 & $0.05\pm 0.008$ & $0.08\pm 0.006$ \\
			071003 & 8.46 & $616\pm57$ & $6.5\pm 0.5$ & - & 1.6 & 3.2 & $0.1\pm 0.006$ & $0.06\pm 0.005$ \\
			050820A & 8.46 &$150\pm31$ & $9.4\pm 1.9$ & 7.35 & 2.61 & 2 & $0.14\pm 0.02$ & $0.1\pm0.02$ \\
			091020 & 8.46 & $399\pm21$ & $10.9\pm 0.6$ & - & 1.71 & 0.45 & $0.08\pm 0.003$ & $0.09\pm0.003$ \\
			000418 & 8.46 & $1085\pm22$ & $18.1\pm 0.3$ & 25 & 1.12 & 0.75 & $0.18\pm 0.003$ & $0.11\pm 0.002$ \\
			100814A & 7.9 & $613\pm23$ & $10.4\pm 0.4$ & - & 1.44 & 0.59 & $0.07\pm 0.002$ & $0.08\pm 0.002$ \\
			980703 & 8.46 & $1370\pm30$ & $10\pm 0.2$ & 7.5 & 0.96 & 0.69 & $0.11\pm0.002$ & $0.08\pm0.002$ \\
			100414A & 8.46 & $524\pm19$ & $8\pm 0.3$ &  - & 1.37 & 7.79 & $0.18\pm0.005$ & $0.09\pm0.004$ \\
			011211 & 8.46 & $162\pm13$ & $13.2\pm 1.6$ & 1.77 & 2.14 & 0.63 & $0.1\pm 0.008$ & $0.13\pm0.006$ \\
			030226 & 8.46 & $171\pm23$ & $6.7\pm 0.1$ & 0.84 & 1.99 & 1.2 & $0.08\pm0.005$ & $0.1\pm0.005$ \\
			990510 & 8.46 & $255\pm34$ & $4.2\pm 0.5$ & 1.2 & 1.62 & 1.8 & $0.07\pm 0.007$ & $0.08\pm0.006$ \\
			991208 & 8.46 & $1804\pm24$ & $7.8\pm 0.1$ & <2.1 & 0.7 & 1.1 & $0.1\pm 0.0009$ & $0.09\pm0.0008$ \\
			970508 & 8.46 & $958\pm11$ & $37.2\pm 0.4$ & 25 & 0.84 & 0.07 & $0.23\pm 0.002$ & $0.23\pm0.002$ \\
			090902B & 8.46 & $84\pm16$ & $14.1\pm 2.6$ & - & 1.82 & 31 & $0.71\pm 0.1$ & $0.22\pm 0.08$ \\
			100418A & 8.46 & $1218\pm12$ & $47.6\pm 0.6$ & 1.11 & 0.62 & 7.8 & $0.47\pm 0.004$ & $0.3\pm0.003$ \\
			090328 & 8.46 & $686\pm26$ & $16.1\pm 0.7$ & 9 & 0.74 & 1 & $0.31\pm0.009$ & $0.2\pm 0.008$ \\
			000911 & 8.46 & $263\pm33$ & $3.1\pm 0.3$ & <1.5 & 1.06 & 8.8 & $0.19\pm 0.02$ & $0.1\pm0.01$ \\
			030329 & 8.46 & $19567\pm28$ & $17.3\pm 0.1$ & 0.55 & 0.17 & 0.18 & $0.1\pm0.0003$ & $0.18\pm 0.0002$ \\
			071010B & 8.46 & $341\pm 41$ & $4.2\pm 0.5$ & - & 0.95 & 0.26 & $0.07\pm 0.007$ & $0.11\pm0.006$ \\
			051022 & 8.46 & $268\pm 32$ & $5.2\pm 0.7$ & - & 0.81 & 6.3 & $0.31\pm0.03$ & $0.16\pm0.02$ \\
			010222 & 8.46 & $93\pm 25$ & $16.8\pm 3.5$ & 0.93 & 1.48 & 13 & $0.42\pm 0.08$ & $0.27\pm0.07$ \\
			981226 & 8.46 & $137\pm 34$ & $8.2\pm 1.6$ & >5 & 1.11 & 0.06 & $0.09\pm0.01$ & $0.21\pm0.01$ \\
			970826 & 8.46 & $144\pm 31$ & $7.8\pm 1.8$ & 2.2 & 0.96 & 3 & $0.32\pm0.05$ & $0.23\pm0.04$ \\
			011121 & 8.7 & $655\pm 40$ & $8.1\pm 0.7$ & 1.3 & 0.36 & 0.46 & $0.25\pm0.01$ & $0.31\pm0.01$ \\
			090424 & 8.46 & $236\pm 37$ & $5.2\pm 0.7$ & 1.6 & 0.54 & 0.45 & $0.22\pm0.02$ & $0.26\pm0.02$ \\
			020405 & 8.46 & $113\pm 17$ & $18.2\pm 3.7$ & 1.67 & 0.69 & 1.1 & $0.58\pm 0.08$ & $0.55\pm0.06$ \\
			020819B & 8.46 & $291\pm 21$ & $12.2\pm 1.1$ & - & 0.41 & 0.08 & $0.22\pm0.01$ & $0.49\pm0.01$ \\
			031203 & 8.46 & $724\pm 19$ & $48\pm 1.3$ & - & 0.11 & 0.001 & $0.27\pm0.006$ & - \\
			\hline  
			\label{tbl:sample}
		\end{tabular}	
	\end{center}
\end{table*}

\section{Interpretation of radio peaks}
\label{sec:interpretation}
We explore here possible interpretations of the observed radio peaks in GRB afterglows within the standard forward shock scenario. Using this framework, there are two possibilities for the nature of these peaks: either it is a jet break transition or a crossing through the observing band by one of the three characteristic synchrotron frequencies, i.e. the self-absorption frequency $\nu_a$, the peak synchrotron frequency $\nu_m$, or the cooling frequency $\nu_c$ (see e.g. \cite{Sari1998,GS2002} for details).

\subsection{Jet Break Time}
The jet break \citep{Rhoads1999,Sari1999,Chevalier2000,Frail2001} takes place when the bulk Lorentz factor decelerates to $\Gamma=\theta_j^{-1}$, where $\theta_j$ is the opening angle of the collimated outflow. The jet break time can be estimated by
\begin{eqnarray}
\label{eq:tjet}
t_j=\!
\left\{ \!
\begin{array}{l}
0.6 (1+z) E_{\rm kin,53}^{1/3} n_0^{-1/3} \theta_{j,0.1}^{8/3} \mbox{days}\quad \mbox{for ISM}\\
0.63 (1+z) E_{\rm kin,53} A_{*}^{-1} \theta_{j,0.1}^4 \quad \mbox{for wind}\\
\end{array} \right.
\end{eqnarray}
where $E_{\rm kin}$ is the isotropic equivalent blast wave energy, $z$ is the redshift, $n_0$ is the density of the circumburst environment if it is homogeneous (in $\mbox{cm}^{-3}$), and $A_* \equiv A/(5\times 10^{11} \mbox{g cm}^{-1})$ is a density parameter in the case that the circumburst medium has a stellar wind structure \citep{Chevalier1999}. An achromatic break is typically seen in the light-curve at $t_j$ \citep[although jet breaks may occur slightly later below the self-absorption break for large viewing angles, see][]{VanEerten2011}, after which the time dependence of the peak flux and characteristic frequencies changes. Specifically, a jet break may be observed as a peak in the radio band in case, for instance, the GRB occurred in an homogeneous medium and $\nu_a<\nu_p<\nu_m$. In such a scenario, the flux would rise as $F_{\nu}(\nu_p)\propto t^{1/2}$ before the jet break and decrease as $F_{\nu}(\nu_p)\propto t^{-1/3}$ afterwards \citep{Sari1999}.

From Equation~\ref{eq:tjet} we see that a rather large opening angle, of order $\theta_j \sim 0.2$ radians, is needed for the jet break time to approach the typical observed peak time $t_p=10$ days. As mentioned in Section~\ref{sec:Sample}, for 23 of the 36 GRBs in our sample there are either measurements or upper limits on the jet break time. In none of these, the observed radio peak is associated with the jet break. In a more general context, \cite{Gao2010} have collected data for 43 bursts for which the jet break time has been estimated in the literature. The average jet break time in this sample is 1.4 days, significantly smaller than the typical time of the radio peaks. Only 5 of the 43 bursts in their sample have $t_j>10$~days. Furthermore, \cite{Ryan2015} have studied a sample of 226 X-ray detected bursts and fitted them with models based on hydrodynamical simulations in order to infer the distribution of opening angles. They find $\theta_0=0.07\pm0.01$, consistent with our canonical values in Equation~\ref{eq:tjet}. Therefore, it is unlikely that most radio peaks are associated with the jet break time. Furthermore, we should expect that in most cases the jet break has already occurred by the time we see the radio peak. Of course, whenever multiple observation bands are available, one can easily determine if the observed peak is due to a jet break.

\subsection{Crossing of a Characteristic Frequency}
As mentioned before, there are three typical frequencies to consider here: $\nu_a$, $\nu_m$, and $\nu_c$. The latter of these can easily be ruled out, as 10~days after the GRB the cooling frequency is expected to reside at $\sim10^{14}-10^{16}$ Hz \citep[e.g.,][]{Beniamini2015,Beniamini2016}. Even with rather extreme microphysical parameters, and accounting for the possible reduction of $\nu_c$ due to Inverse Compton cooling, $\nu_c$ should be larger than $\sim10^{12}$~Hz, which is still at least two orders of magnitude larger than the radio frequencies we are focusing on here. Furthermore, for $t>t_j$, $\nu_c$ stops decreasing as a function of time \citep{Sari1999,DeColle2012,VanEerten2013}, so even at very late times one does not expect to see the cooling break pass through the radio band.

\begin{figure*}
	\centering
	\includegraphics[scale=0.35]{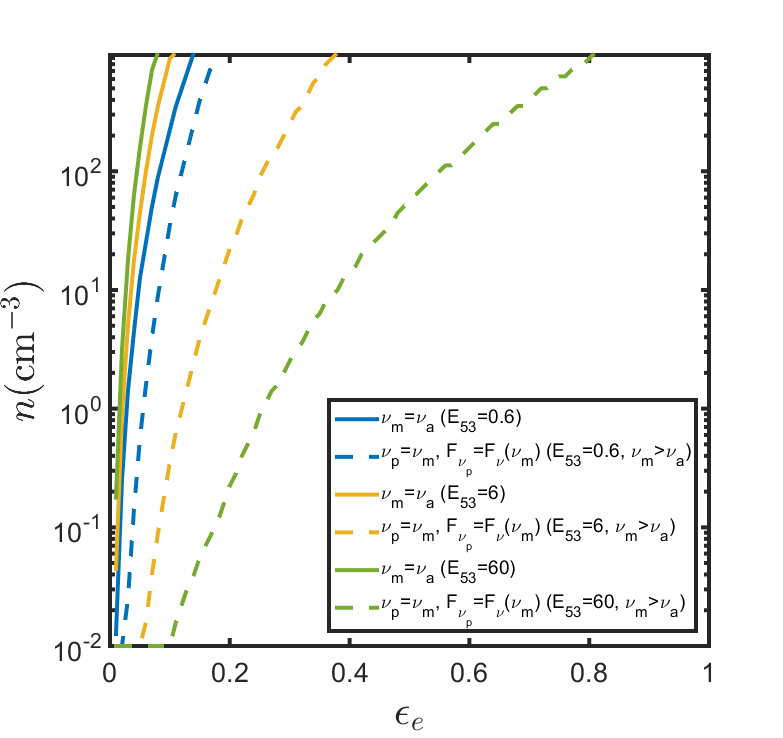}
	\includegraphics[scale=0.35]{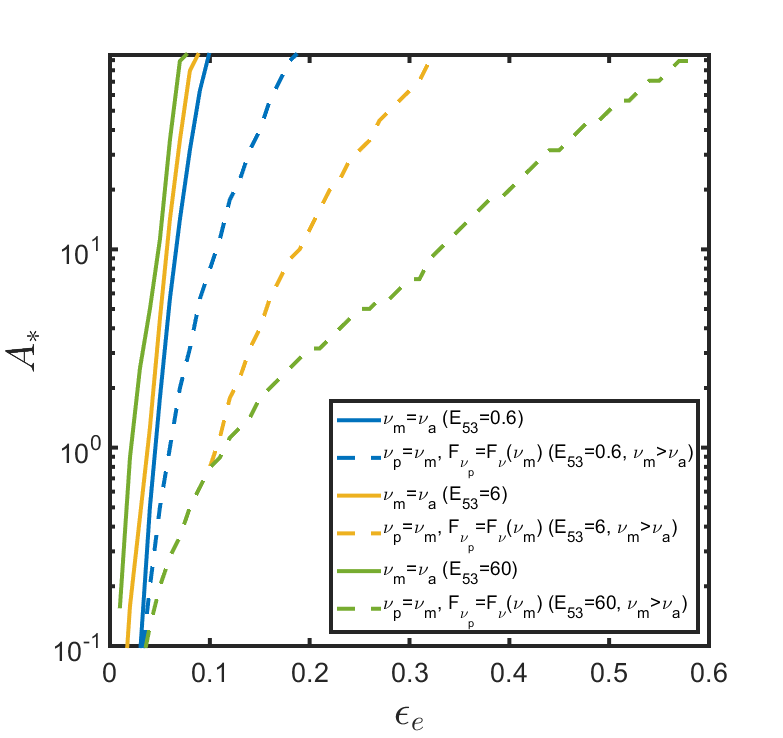}
	\caption
	{\small Dashed lines depict $n(\epsilon_e)$ (for a homogeneous medium, left panel) and $A_*(\epsilon_e)$ (for stellar wind medium, right panel) implied by the association of the radio peak with the transition of $\nu_m$ through the radio band. Different colours depict different blast-wave energies. We consider an order of magnitude spread around the average value in the sample $\langle E_{\rm kin} \rangle = 6.4 \times 10^{53}$ergs; see Section~\ref{sec:interpretation} for details. We assume here a typical jet opening angle of $\theta_j=0.1$; the calculated frequencies and fluxes take into account post jet break steepening when necessary. Solid lines are the upper limits for the consistency requirement $\nu_m>\nu_a$; The allowed region lies below and to the right of these curves. In all cases shown here, the curves implied by the radio peaks are entirely within the self-consistent regime.}
	\label{fig:numconsistency}
\end{figure*}

\begin{figure*}
	\centering
	\includegraphics[scale=0.35]{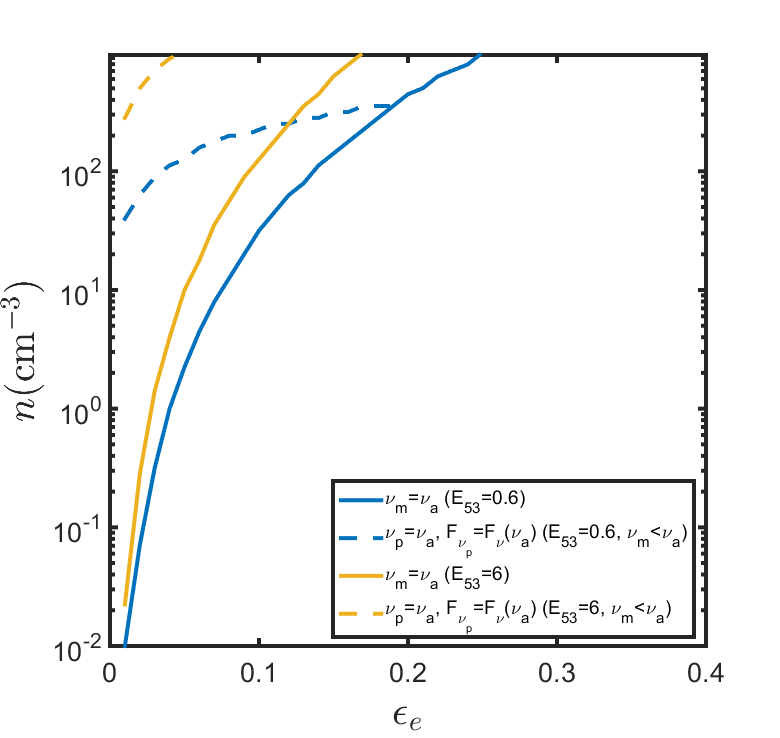}
	\includegraphics[scale=0.35]{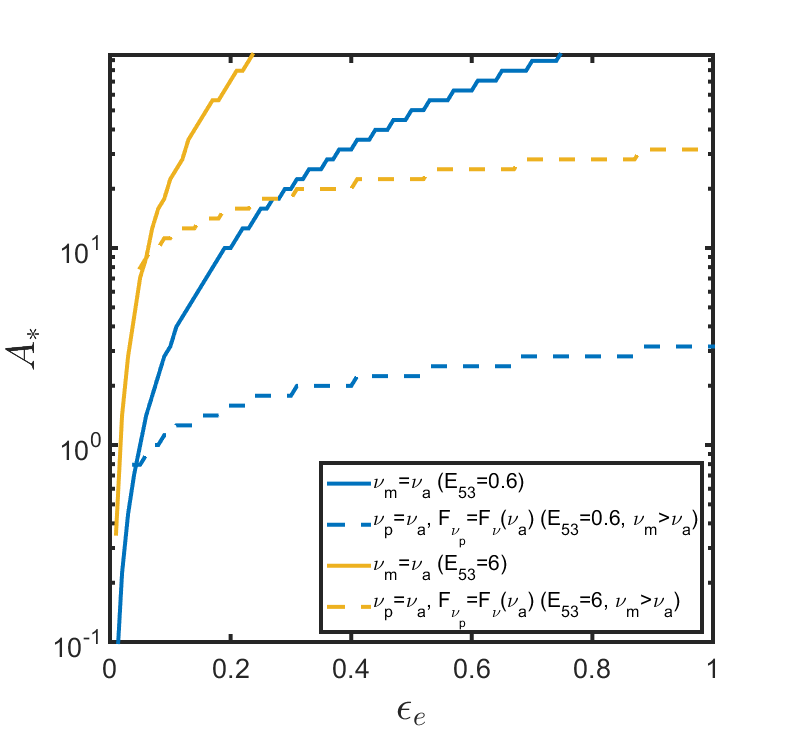}
	\caption
	{\small Dashed lines depict $n(\epsilon_e)$ (for a homogeneous medium, left panel) and $A_*(\epsilon_e)$ (for stellar wind medium, right panel) implied by the association of the radio peak with the transition of $\nu_a$ through the radio band. Different colours depict different blast-wave energies. We consider an order of magnitude spread around the average value in the sample $\langle E_{\rm kin} \rangle = 6.4 \times 10^{53}$ergs; see Section~\ref{sec:interpretation} for details. We assume here a typical jet opening angle of $\theta_j=0.1$; the calculated frequencies and fluxes take into account post jet break steepening when necessary. For a homogeneous medium, solid lines are the lower limits for the consistency requirement $\nu_a>\nu_m$; The allowed region lies to the top and left of these curves. For a wind medium solutions are only found with $\nu_a<\nu_m$. In this case the solid lines are the upper limit on this consistency requirement and allowed solutions are to the right and bottom of these curves. For a homogeneous medium, allowed solutions require $n\gtrsim 60 E_{\rm kin,53}\mbox{ cm}^{-3}$ and $\epsilon_e \lesssim 0.2$, whereas for a wind medium, $A_* \gtrsim 3 E_{\rm kin,53}$ and $\epsilon_e \gtrsim 0.05$ is required. }
	\label{fig:nuaconsistency}
\end{figure*}

We are thus left with two options, $\nu_a$ or $\nu_m$, for the observed peak frequency $\nu_p$. The possibilities are further constrained by the requirement that the transition should be observed as a peak in the radio (so the radio flux should increase before and decrease after the transition). Given the typical time dependencies of $\nu_a$ or $\nu_m$ in the forward shock model, this means that if $\nu_p=\nu_m$, then we must have $\nu_m>\nu_a$, whereas if $\nu_p=\nu_a$ then either $\nu_a>\nu_m$ or $\nu_a<\nu_m$ within the connection that it is a stellar wind medium and $t_p>t_j$ (with $t_p$ the observed peak time). The location of these frequencies and their fluxes are determined by four physical parameters: $\epsilon_e$, $\epsilon_B$, $E_{\rm kin}$, and $n$ or $A_*$ (depending on the structure of the circumburst medium). A fifth parameter, the index of the electrons' energy spectrum injected at the shock, $p$, is also strictly required to fully determine the observed parameters, but is more restricted ($2.2 \lesssim p\lesssim 2.8$) and does not hugely affect the observables. Furthermore, in cases when the jet break time is not measured, one needs additional knowledge of the jet's opening angle. Associating the typical observed frequency ($\nu_p=10$ Ghz) and peak flux ($F_{\nu}(\nu_p)=1$ mJy) with those of the crossing frequency at $t_p\approx 10$ days (accounting for post jet break steepening when required) provides us with two equations relating the micro-physical parameters. Since $\epsilon_B$ is likely the least constrained parameter \citep{Lemoine2013,Wang2013,BD2014,Santana2014,Zhang2015,Beniamini2015,Beniamini2016}, we use these equations to remove their dependence on this parameter and to obtain solutions for $n(E_{\rm kin},\epsilon_e)$ or $A_*(E_{\rm kin},\epsilon_e)$ for a given association of the peak frequency and flux. We can further constrain the blast wave isotropic equivalent energy $E_{\rm kin}$, by knowledge of the isotropic equivalent energy released in gamma-rays during the prompt phase, $E_{\gamma,\rm iso}$. Studies of GRB afterglows in the optical, X-rays and GeV gamma-rays suggest that the efficiency of the prompt phase, defined as $\epsilon_{\gamma}\equiv E_{\gamma,\rm iso} / (E_{\rm kin}+E_{\gamma,\rm iso})$ is $\sim 0.15^{+0.08}_{-0.05}$ \citep{Fan2006,Beniamini2015,Beniamini2016}. Since this parameter varies much less from burst to burst than $E_{\rm kin}$ \citep{Nava2014,Beniamini2016}, it can effectively remove the dependence on $E_{\rm kin}$ in the aforementioned equations. If we assume a constant $\epsilon_{\gamma}=0.15$, the (logarithmic) average of $E_{\rm kin}$ for the bursts in the sample is $\langle E_{\rm kin} \rangle = 6.4 \times 10^{53}$ergs and its standard deviation is 0.85 orders of magnitude. We use these as guidelines for the choice of $E_{\rm kin}$ in Figures~\ref{fig:numconsistency} and~\ref{fig:nuaconsistency}. We point out that for the main analysis presented in Section~\ref{sec:implications} we do not require any assumptions on the unknown $\epsilon_{\gamma}$ (or other physical parameters') distribution. In fact, as will be discussed in Section~\ref{sec:discussion}, any scatter in $\epsilon_{\gamma}$, or any of the other physical parameters, would only strengthen our conclusions regarding the required narrow scatter of $\epsilon_e$ implied by the analysis presented here.

We consider first (Figure~\ref{fig:numconsistency}) the case in which the peak frequency is due to the crossing of $\nu_m$ through the observed radio band. Dashed lines depict the required values of $n(\epsilon_e)$ (left panel) or $A_*(\epsilon_e)$ (right panel).
In order to check that these solutions are consistent, we verify that they can result in $\nu_m>\nu_a$, given a range of values for $E_{\rm kin}$ between $6\times10^{52}$ and $6\times10^{54}$~ergs, and a reasonable value for the unknown $\epsilon_B$ (which we conservatively assume here lies in the range $10^{-6}\lesssim \epsilon_B \lesssim 1$).
The solid curves in Figure~\ref{fig:numconsistency} depict the upper limit for the consistency requirement $\nu_m>\nu_a$ (corresponding to $\nu_m=\nu_a$ with $\epsilon_B=1$); self-consistent solutions would lie below and to the right of these curves.
We show in Figure~\ref{fig:numconsistency} that in all considered cases the implied physical parameters are always consistent with the necessary assumption that $\nu_m>\nu_a$.

The alternate possibility, that the radio peak is the crossing of $\nu_a$, is considered in Figure~\ref{fig:nuaconsistency}. As before, dashed lines depict the requirements on $n(\epsilon_e)$ and $A_*(\epsilon_e)$.
In the case of a homogeneous medium, allowed solutions require a large density $n\gtrsim 60 E_{\rm kin,53}\mbox{ cm}^{-3}$. The solid curves in the left panel Figure~\ref{fig:nuaconsistency} depict the lower limit for the consistency requirement $\nu_a>\nu_m$ (corresponding to $\nu_m=\nu_a$ with $\epsilon_B=10^{-6}$); self-consistent solutions would lie to the top left of these curves, leading to the requirement $\epsilon_e \lesssim 0.2$. Therefore, part of parameter space allows for the possibility of $\nu_a>\nu_m$, but given the strict requirements on the physical parameters in this case, it seems considerably less likely that the observed peaks correspond to the transition of $\nu_a$.
For a wind medium, even when $\nu_a<\nu_m$, the crossing of $\nu_a$ would still be seen as a radio peak. In fact, no solutions with $\nu_a>\nu_m$ are obtained in this case. The association of $\nu_a$ with these peaks translates to a requirement of a large wind parameter $A_* \gtrsim 3 E_{\rm kin,53}$. As in Figure~\ref{fig:numconsistency}, the solid curves in the right panel of Figure~\ref{fig:nuaconsistency} depict the upper limit for the consistency requirement $\nu_m>\nu_a$, which leads to a lower limit on $\epsilon_e$ in this case of $\epsilon_e \gtrsim 0.05$.
As for the case of the homogeneous medium, we conclude that given the allowed parameter space for such solutions, it is unlikely that the observed peaks would correspond to the transition of $\nu_a$.
Of course, in cases where some of the microphysical parameters are better constrained by additional observations, or alternatively, whenever the rise and decay rate (or the spectra) before and after the peak are available, one can identify more clearly which characteristic peak is crossing through the observed band.

We conclude that the radio peak is most likely to be the transition of $\nu_m$ through the observed band. In the following we will consider the implication of this association for our sample of GRBs. We stress that even in case this association is wrong for part of the GRBs in the sample, this will actually broaden the measured spread of $\epsilon_e$ discussed in Section~\ref{sec:implications}, and will imply that the intrinsic spread is even narrower than the one we find.

\section{Implication of radio peaks for the parameter distributions}
\label{sec:implications}
Identifying the radio peak with the transition of $\nu_m$ through the band, i.e. $\nu_p=\nu_m$ and $F_{\nu_p}=F_{\nu}(\nu_m)$, and using the equations for the synchrotron frequency and flux from \cite{GS2002}, adding corrections to the post jet break flux and frequency when necessary, allows us to compute the following parameters
\begin{equation}
\label{eq:Psi1}
\begin{split}
&\Psi_{\rm ISM}\!\equiv \!\bigg(\frac{261.4(1\!+\!z)^{1/2} \nu_p t_p^{3/2}E_{\gamma,\rm iso, 53}^{1/2}}{ 10^{15}d_{28}^2 F_{\nu_p} \mbox{ max}(1,t_p/t_j)^{1/2}}\bigg)^{1/2}\!\\
&=\!\bigg(\frac{p\!-\!0.67}{p\!+\!0.14}\bigg)^{1/2}\frac{\epsilon_e(p\!-\!2)}{0.177(p\!-\!1)} \bigg(\frac{1\!-\!\epsilon_{\gamma}}{\epsilon_{\gamma}}\bigg)^{-1/4}\! n_0^{-1/4}
\end{split}
\end{equation}
\begin{equation}
\label{eq:Psi2}
\begin{split}
&\Psi_{\rm wind}\!\equiv \! \bigg(\frac{249.4 (1\!+\!z) \nu_p t_p}{ 10^{15}d_{28}^2 F_{\nu_p} }\bigg)^{1/2}\!\\
&=\!\bigg(\frac{p\!-\!0.69}{p\!+\!0.12}\bigg)^{1/2}\frac{\epsilon_e(p\!-\!2)}{0.277(p\!-\!1)} A_*^{-1/2}
\end{split}
\end{equation}
where $\nu_p$ is in Hz, $F_{\nu_p}$ is in mJy, $t_p$ is in days and $d_{28}$ is the luminosity distance in units of $10^{28}$cm. The left hand side (top lines) of these equations can be computed directly from observations, assuming that $t_j$ can be measured (or that a lower limit on $t_j$ larger than $t_p$ can be obtained). This is indeed the case for $64\%$ of the bursts in the sample. Notice that for the wind scenario the result is independent of $t_j$, because the time dependence of $F_{\nu}(\nu_m)$ and $\nu_m$ steepen by the same amount after the jet break, and their ratio is thus unchanged. When $t_j$ measurements are not available, one can estimate the jet break time for a given $\theta_j$ according to Equation~\ref{eq:tjet}. In what follows we canonically assume $\theta_j=0.1$. Interestingly, the right hand side (bottom lines) of Equations~\ref{eq:Psi1} and~\ref{eq:Psi2} is linear in $\epsilon_e$ and weakly dependent on $\epsilon_{\gamma}$, $n_0$, $A_*$, or $p$. The numerical factor in the denominator is chosen such that it cancels the $p,\epsilon_{\gamma}$ dependencies for $p=2.5$ and $\epsilon_{\gamma}=0.15$. With this notation, $\Psi=\epsilon_e$ for $p=2.5, \epsilon_{\gamma}=0.15, n_0=1, A_*=1$. Thus, measurements of radio peaks in GRB afterglows can be used for estimating $\epsilon_e$. We note that even for those cases when $t_j$ is before the radio peak, yet cannot be measured, $\Psi$ still depends mainly on $\epsilon_e$ and scales as $\Psi_{\rm ISM} \propto \epsilon_e \epsilon_{\gamma}^{1/3} n_0^{-1/6} \theta_{j,-1}^{-2/3}$, omitting here the weak $p$ dependence for clarity, and assuming $\epsilon_{\gamma}\ll1$); for a wind medium there is no change in this case.

\begin{figure*}
	\centering
	\includegraphics[scale=0.39]{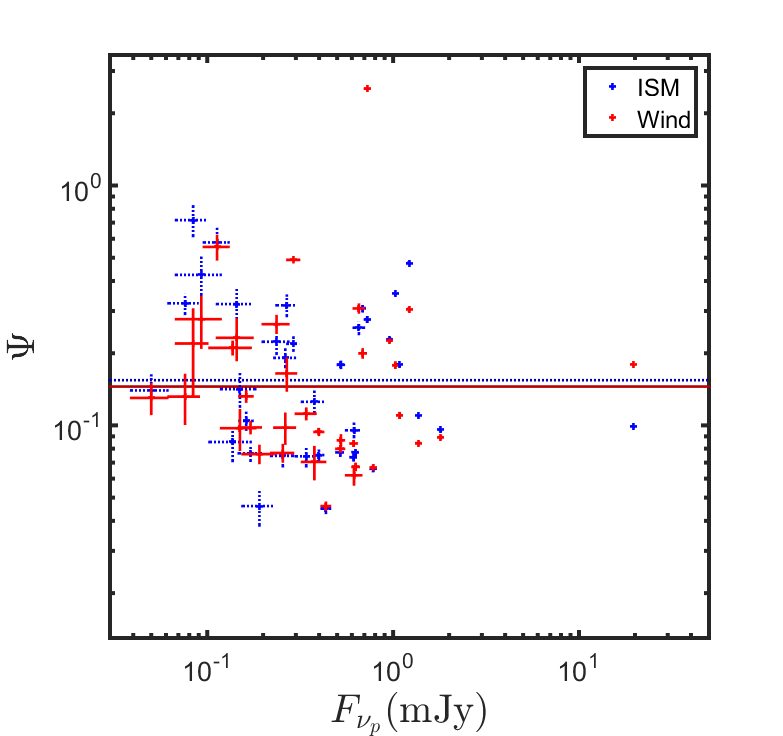}
	\includegraphics[scale=0.39]{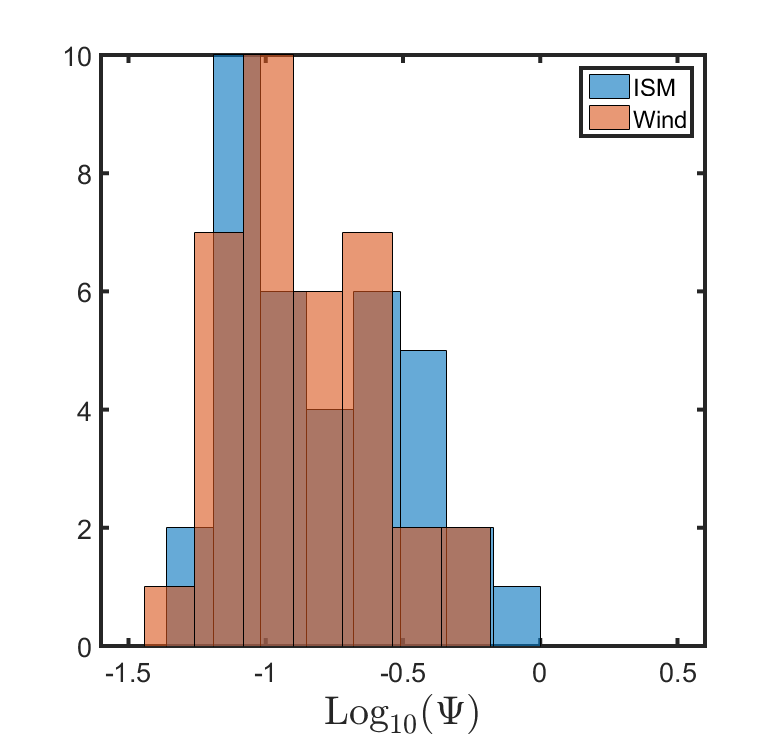}
	\caption
	{\small Distribution of $\Psi$ (see Equations~\ref{eq:Psi1} and~\ref{eq:Psi2} for a definition) for the 36 GRBs in our radio peak sample. The left panel shows $\Psi$ as a function of the radio peak flux; the right panel shows the number distribution. Results are shown for both a homogeneous (blue, dotted lines) and a stellar wind medium (red, solid lines). Average values (in log-space) are denoted by horizontal lines in the left panel for both environments. Assuming that the observed peak is the transition of the synchrotron frequency $\nu_m$ through the observing band, then $\Psi\approx \epsilon_e$ (with equality for $p=2.5$, $\epsilon_{\gamma}=0.15$, $n_0=1$, $A_*=1$, see Section~\ref{sec:implications}).}
	\label{fig:Psidist}
\end{figure*}

In Table~\ref{tbl:sample} we list the observed parameters for each burst along with the derived values of $\Psi_{\rm ISM},\Psi_{\rm wind}$. Errors are propagated from the errors in the fluxes and times of the peaks. The distributions of $\Psi_{\rm ISM},\Psi_{\rm wind}$ are plotted in Figure~\ref{fig:Psidist} as a function of the peak radio flux, which is the main criterion for the detectability of the radio peaks. The lack of correlation between $\Psi$ and $F_{\nu_p}$ indicates that the resulting distribution of $\Psi$ does not suffer from a selection bias and can be extended to the general GRB population (a discussion of the optimal conditions for searching for such peaks in future observations is given in Section \ref{sec:future}). We find that the average (in log-space) values of $\Psi$ are $\langle \Psi_{\rm ISM} \rangle=0.15$ and $\langle \Psi_{\rm wind} \rangle=0.13$. We restate here for $p=2.5$, $\epsilon_{\gamma}=0.15$, $n_0=1$, $A_*=1$, and $\theta_j=0.1$ (when $t_j$ is not measured directly), $\Psi=\epsilon_e$ and that $\Psi$ depends rather weakly on these other parameters. Indeed, the average value of $\Psi$ matches remarkably well the typical values of $\epsilon_e$ expected both from theory \citep{Sironi2011}, detailed modelling of GRB afterglows \citep{Santana2014}, and clustering of GeV gamma-ray light-curves \citep{Nava2014}. This can be viewed as further justification for the association of the peaks with the transition of $\nu_m$ through the radio band, but we have also verified that the obtained values of $\epsilon_e$ are consistent with the assumption $\nu_m>\nu_a$. This self-consistency test is passed by all of the GRBs in the sample. In one burst, GRB 031203, if the explosion took place in a wind medium, the jet would have most likely become non-relativistic before $t_p$ \citep{Soderberg2004}, unless the wind parameter in this burst was extremely small ($A_* \lesssim 4\times 10^{-3}$). Following our selection criteria of Section~\ref{sec:Sample}, we therefore do not include this burst in the analysis for the wind medium case.

The measured scatter of $\Psi$ can be translated to the scatter of the physical parameters. Using Equations~\ref{eq:Psi1} and~\ref{eq:Psi2}, we find that (assuming no inter-correlation between the different burst parameters):
\begin{equation}
\label{eq:scatterPsiISM}
\sigma_{\log \Psi_{ISM}}^2\!=\!\left\{ \!
\begin{array}{l}
\sigma_{\log \epsilon_e}^2+ \frac{1}{16}\sigma_{\log \epsilon_{\gamma}}^2+ \frac{1}{16}\sigma_{\log n_0}^2\quad\quad\quad t_j \mbox{  known}\\
\sigma_{\log \epsilon_e}^2\!+\! \frac{1}{9}\sigma_{\log \epsilon_{\gamma}}^2\!+\! \frac{1}{36}\sigma_{\log n_0}^2 \!+\! \frac{4}{9}\sigma_{\log \theta_{j,-1}}^2 \quad \mbox{else}\\
\end{array} \right.
\end{equation}
\begin{equation}
\label{eq:scatterPsiwind}
\sigma_{\log \Psi_{wind}}^2=\sigma_{\log \epsilon_e}^2+ \frac{1}{4}\sigma_{\log A_*}^2
\end{equation}
Based on the distributions of $\Psi_{\rm ISM}$ and $\Psi_{\rm wind}$ shown in Figure~\ref{fig:Psidist}, we find that $\sigma_{\log_{10} \Psi_{ISM}}^2=0.1$ and $\sigma_{\log_{10} \Psi_{wind}}^2=0.07$. Using Equations~\ref{eq:scatterPsiISM} and~\ref{eq:scatterPsiwind}, these values provide strong upper limits on the scatter of $\epsilon_e$. We find that the scatter in this parameter can be $\sigma_{\log_{10} \epsilon_e}<0.31$ orders of magnitude at most for a homogeneous medium, or $\sigma_{\log_{10} \epsilon_e}<0.26$ orders of magnitude for a stellar wind medium. Any scatter in $p$, $\epsilon_{\gamma}$, $n_0$, or $A_*$, which are of course expected, would reduce the upper limits on the scatter of $\epsilon_e$. In Section~\ref{sec:discussion} we discuss that the scatter of $\Psi$ found in this analysis is already in contention with the distributions of GRB parameters suggested by various broadband modeling efforts.

\section{Future observations}
\label{sec:future}
In this section we reverse the logic applied in the rest of this paper: instead of trying to deduce the distributions of physical parameters from observations, we consider some typical ranges for the intrinsic burst parameters (implied by the current work and others in the literature), and speculate on the detectability of radio peaks associated with the crossing of $\nu_m$ in different radio bands. We apply a Monte Carlo method: we assign a given distribution to each free parameter, draw $10^3$ random values according to those distributions, and using the forward shock afterglow synchrotron model (with corrections for jet breaks when applicable), calculate the time $t_p$ at which $\nu_p=\nu_m$, and the spectral flux $F_{\nu_p}= F_{\nu}(\nu_p, t_p)$; and we discuss the detectability of those peaks as a function of $\nu_p$.

We list here the parameter distributions assumed in this section, and we only explore a homogeneous circumburst medium. We use the studies of GRBs detected with the {\it Fermi} Gamma-ray Burst Monitor, to estimate the distribution of  $E_{\gamma,\rm iso}$ and $z$ \citep{Goldstein2016}. Following that paper, we assume a log-normal distribution peaking at $\langle E_{\gamma, \rm iso} \rangle=1.4\times 10^{53}$erg with a standard deviation of $\sigma_{\log_{10} E_{\gamma, \rm iso}}=0.84$. The redshift distribution ranges between $1\lesssim z \lesssim 3$ \citep[see][for the exact functional form]{Goldstein2016}. We adopt a log-normal distribution for $\epsilon_{\gamma}$ with $\langle \epsilon_{\gamma} \rangle=0.15$ and $\sigma_{\log_{10} \epsilon_{\gamma}}=0.2$ \citep{Nava2014,Beniamini2015,Beniamini2016}. Following the results found in the present work, and in former studies \citep{Nava2014,Santana2014}, we assume a log-normal distribution for $\epsilon_e$, with $\langle \epsilon_e \rangle=0.15$ and $\sigma_{\log_{10} \epsilon_e}=0.2$. $\epsilon_B$ and $n_0$ are not as well constrained as the other parameters (see also Section~\ref{sec:discussion}). Here we adopt log-normal distributions with $\langle \epsilon_B \rangle = 10^{-4}$, $\sigma_{\log_{10} \epsilon_B}=1$, $\langle n_0 \rangle =1$ and $\sigma_{\log_{10} n_0}=1$ \citep[see also][]{Santana2014,Zhang2015,Beniamini2016}). Furthermore, we consider a uniform distribution of $p$ in the range $2.2\leq p \leq 2.8$ (see Section~\ref{sec:discussion}). Finally, we take a log-normal distribution for $\theta_j$ with $\langle \theta_j \rangle=0.1$ and $\sigma_{\log_{10} \theta_j}=0.2$ \citep{Goldstein2016}.

\begin{figure*}
	\centering
	\includegraphics[scale=0.39]{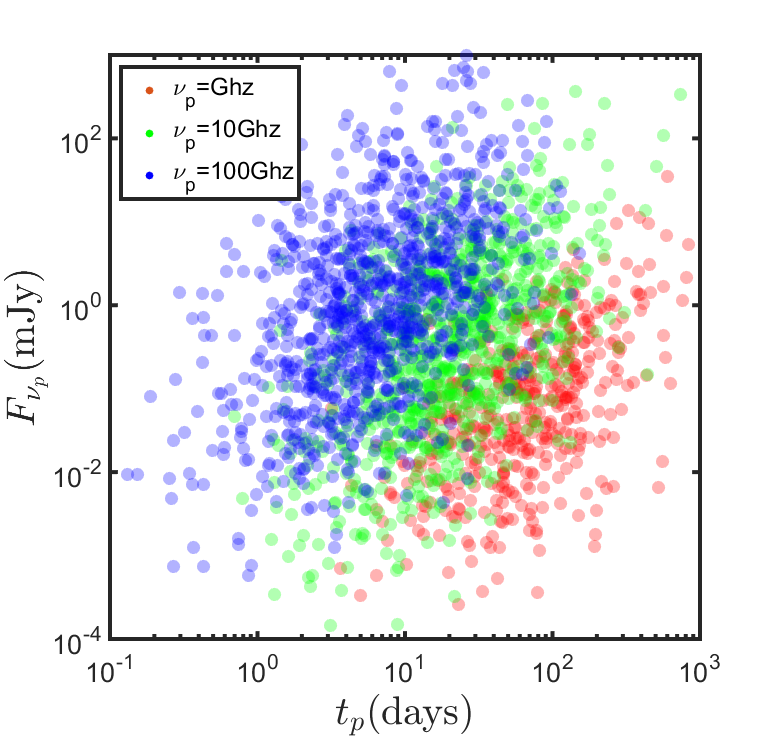}
	\caption
	{\small Distribution of times and fluxes of radio peaks associated with the transition of $\nu_m$ through the observed frequency $\nu_p$ for a sample of simulated bursts (see Section~\ref{sec:future} for details). The fraction of bursts for which $\nu_m>\nu_a$ when $\nu_m$ crosses $\nu_p$ (as required to cause a radio peak, see Section~\ref{sec:interpretation}) and $\Gamma(t_p)>2$ is $45\%$ for $\nu_p=1$~GHz, $88\%$ for $\nu_p=10$~GHz and $99\%$ for $\nu_p=100$~GHz.}
	\label{fig:future}
\end{figure*}

The distribution of $t_p$ versus $F_{\nu_p}$ for different $\nu_p$ is shown in Figure~\ref{fig:future}. For each simulated burst we check whether $\nu_m>\nu_a$ when $\nu_m$ crosses the observed band (as required in order for this transition to cause a radio peak, see Section~\ref{sec:interpretation}), and that $\Gamma(t_p)>2$. We find that for $\nu_p=1$~GHz only $45\%$ of simulated bursts satisfy these criteria, since at this band, $\nu_a$ is often larger than $\nu_m$. This fraction increases to $88\%$ for $\nu_p=10$~GHz and reaches $99\%$ at $\nu_p=100$~GHz. Although the exact fractions would of course depend on the assumptions regarding the underlying distributions, the general trend is that the viability for $\nu_m>\nu_a$ goes up with frequency and requires $\nu_p\gtrsim10$~GHz for the vast majority of the radio peaks to be dominated by the $\nu_m$ transition. On average, $t_p=18(\nu_p/10~\rm{GHz})^{-1/2}$ days, $F_{\nu_p}=0.3(\nu_p/1~\rm{GHz})^{1/2-2/3}$ mJy, with $\nu_p$ in GHz. Thus, at larger frequencies these peaks would be more easily detectable. Assuming that bursts can be detected in the radio as early as $\sim0.3$~days after the GRB trigger, we suggest that observing the $10-100$~GHz band, between $0.3-30$~days after the GRB, would yield ideal conditions for future detections of peaks associated with the transition of $\nu_m$. Observations at lower radio frequencies ($\sim1$~GHz) would be very useful to determine the location of the self-absorption frequency, to help in determining a larger set of physical parameters in broadband modeling.

\section{Discussion}
\label{sec:discussion}
Different authors have performed broadband modeling studies on samples of GRB afterglows including radio data \citep[e.g.,][]{WijersGalama1999,PK2001,PK2002,Yost2003,Cenko2010,Cenko2011} assuming that the radiation is synchrotron emission produced by electrons accelerated at the forward shock. These groups find widely varying values for the model parameters ($\epsilon_e$, $\epsilon_B$, $E_{\rm kin}$, $p$, $n_0$, $A_*$) as well as different results for the same GRBs.
An illuminating example, is the case of GRB 970508. This GRB has been widely studied by many groups, which inferred mutually contradicting and widely varying values of the intrinsic parameters. Some studies suggest values of $E_{\rm kin}$ as low as $5\times 10^{50}$ergs \cite{Frail2000}, while others suggest values as large as $3.7\times 10^{52}$ergs \cite{Yost2003}. The external density is also highly unconstrained. While most authors find better fits with an ISM environment, \cite{Chevalier2000,PK2002} find better fits with  wind medium. The typical densities vary from $n=0.03 \mbox{cm}^{-3}$ \cite{WijersGalama1999} to $n=7.465 \mbox{cm}^{-3}$ \cite{Leventis2013}. Finally, the values of $\epsilon_B,\epsilon_e$ in the latter two studies also vary significantly, from $\epsilon_B=0.089,\epsilon_e=0.12$ to $\epsilon_B=0.55,\epsilon_e=0.595$. These major differences (which are also significantly larger than the reported uncertainties in the values of the derived parameters above), clearly demonstrate the intrinsic complexities involved in broadband studies.

Typical ranges found for the physical parameters in broadband modeling studies are \citep[summarized in the review by][]{GvdH2014}: $10^{-5}\lesssim \epsilon_B \lesssim 10^{-1}$, $10^{50}\mbox{ ergs}\lesssim E_{\rm jet} \lesssim 3\times 10^{52}\mbox{ ergs}$, $2\lesssim p \lesssim 3$, $10^{-3}\lesssim n_0 \lesssim 10^2\mbox{ cm}^{-3}$, $10^{-2} \lesssim A_* \lesssim 10^{-0.5}$ (where $E_{\rm jet}$ is the true, collimated corrected rather than isotropic equivalent, blast wave energy). Most important in the context of the present work, these studies find $5\times 10^{-3} \lesssim \epsilon_e \lesssim 3\times 10^{-1}$ with $\sigma_{\log_{10} \epsilon_e, \rm lit}=0.49$ for the sample of bursts in \cite{Cenko2010,Cenko2011} and $\sigma_{\log_{10} \epsilon_e, \rm lit}=0.47$ for the bursts studied by \cite{PK2002}. In both cases $\sigma_{\log_{10} \epsilon_e, \rm lit}>\sigma_{\log_{10} \Psi}\approx 0.3$ where the right hand side is an upper limit on $\sigma_{\log_{10} \epsilon_e}$ as implied by radio peaks (see Section \ref{sec:implications}). The situation becomes even worse when taking into account the enormous spread of homogeneous densities found in those studies (for wind there is an insufficient number of well modelled GRBs to calculate a reliable spread): the values reported in those studies have a spread $\sigma_{\log_{10} n_0}=1.55$. Plugging this value into Equation \ref{eq:scatterPsiISM}, we see that this spread alone would have accounted for slightly more than the measured value of $\sigma_{\log_{10} \Psi_{ISM}}$. This would imply a negligible spread in $\epsilon_e$ (and $\epsilon_{\gamma}$) in order for the densities found in those studies to be consisted with our findings, and will make the $\sigma_{\log_{10} \epsilon_e}$ discrepancy even more severe. Inter-correlations between the physical parameters could, in principle, help to resolve this apparent contradiction. However, due to the weak dependence of $\Psi$ on $n_0$, such correlations would have to be rather strong in order to affect our conclusions here, and there is no obvious reason to expect the existence of such strong correlations (nor are they inferred from broadband modeling studies).

The range of $\epsilon_e$ values reported in broadband modeling studies is also in contradiction with limits from the clustering of GeV light-curves \citep{Nava2014} that suggest $\sigma_{\log_{10} \epsilon_e}\leq 0.19$, in agreement with our results from the radio peaks. The fact that the typical values and scatter of $\epsilon_e$ found from both extremes of the electromagnetic spectrum (radio and GeV) are mutually consistent, lends further credence to the results found in either method individually. Furthermore, since these methods used data collected at different times, the implication is that $\epsilon_e$ remains constant throughout the evolution of the blast-wave, and is not strongly dependent on the blast wave's Lorentz factor \citep{Eichler2005,Sironi2011,Warren2015}.

The scatter of the physical parameters, coupled with the fact that broadband modeling of the same bursts lead to different implied $\epsilon_e$ values (as described above), highlights the difficulty of trying to obtain physical parameters based on broadband modeling. Two main reasons could potentially lead to these discrepancies: (i) there is some missing ingredient in the physical model that affects the multi-wavelength spectra and light-curves, and/or (ii) possible degeneracies between the model parameters when insufficient observational data are available. Examples for the first of these reasons include contributions due to the reverse shock  \citep{Uhm2007,Genet2007,BM2017}, suppression of the (typically X-ray) synchrotron flux due to Inverse Compton cooling \citep{SariEsin2001,Zou2009,Beniamini2015} and viewing angle effects \citep{EichlerGranot2006,VanEerten2010,DeColle2012}. These complications demonstrate the need for robust diagnostic tools using one or two observables, as presented in this paper, that could help to pin down a given parameter in a relatively clean manner with as little dependencies on other physical parameters as possible.

This means that diagnostic tools like the one presented in this paper \citep[and also, for instance, in][]{Nava2014,BD2014,Beniamini2015}, can be used together with broadband modeling techniques to further pin down the physical processes at play in a given GRB. An example of this is a GRB that was not part of our sample, because it occurred after publication of \citet{CF2012}, but is one of the most well-studied GRBs to date: GRB\,130427A. This GRB had very well sampled light curves all across the electromagnetic spectrum, including the radio regime \citep{Laskar2013,Perley2014,Anderson2014,VanDerHorst2014}. The radio light curves and spectra of this GRB showed a complex structure, strongly suggesting two emission components. Based on the 15~GHz light curve \citep{Anderson2014,VanDerHorst2014}, there were two peaks at different times and flux levels, and from those measured values we estimate the following values for $\Psi$: $\Psi_{\rm ISM}\sim0.07$ and $\Psi_{\rm wind}\sim0.05$ for the first peak, while $\Psi_{\rm ISM}\sim6.6$ and $\Psi_{\rm wind}\sim1.7$ for the second peak. The first conclusion that one can draw is that the $\Psi$ values for both peaks are very different, suggesting different physical interpretations of those peaks. This agrees well with the interpretation that the first peak is emission from the reverse shock, while the second peak is emission from the forward shock. Consequently, one could apparently conclude that GRB\,130427A is an outlier for the narrow distribution of $\epsilon_e$ presented in this paper. However, broadband modeling of GRB\,130427A by all of the aforementioned studies has shown that the density of its circumburst medium is extremely low, by about 3 to 4 orders of magnitude, and that this medium is structured like a stellar wind. Since $\Psi$ is proportional to $A_*^{-1/2}$, this puts $\epsilon_e$ right in its narrow distribution. This is a great example of how the method presented in our paper can be used as a powerful diagnostic tool, and as a consistency check for broadband modeling of GRB afterglows. It also shows that although it is a great tool, broadband modeling including observations from low radio frequencies to high energy $\gamma$-rays are necessary to completely pin down the physics of GRB afterglows.

\section{Conclusions}
\label{sec:conclusions}

In this paper we have presented a new method to constrain physical parameters related to GRB afterglows, based on their radio light curve peaks. By combining the equations for peak flux and peak frequency in such a way that the highly uncertain parameter $\epsilon_B$ disappears, we construct a parameter $\Psi$ which is proportional to $\epsilon_e$ and weakly dependent on other physical parameters, such as the blast wave energy and the density of the circumburst medium. We find that $\langle \Psi_{\rm ISM} \rangle=0.15$ for a homogeneous medium and $\langle \Psi_{\rm wind} \rangle=0.13$ for a stellar wind medium. These $\Psi$ values are exactly equal to $\epsilon_e$ for $p=2.5$, $\epsilon_{\gamma}=0.15$, $n_0=1$, $A_*=1$, and $\theta_j=0.1$. The average value of $\Psi$ matches remarkably well with the typical values of $\epsilon_e$ expected both from theory and clustering of GeV gamma-ray light-curves. Based on our study, we also constrain the spread in parameter values, and find that the large spreads found in broadband modeling studies for the energy, density, and microphysical parameters, are inconsistent with our analysis of the radio peaks. Assuming that the method presented here is robust, this result may be caused by different groups using different modeling codes with different input assumptions, or insufficient available data for some GRBs. Another potentially important factor is that there might be some missing ingredients in the physical model, such as contributions from the reverse shock, suppression of the synchrotron flux due to Inverse Compton cooling, and viewing angle effects. Our method for the analysis of radio light curve peaks is suggested as a diagnostic tool that can be used together with broadband modeling codes, to get a better handle on all the physical processes at play in GRB afterglows. The recent upgrade of several radio observatories, and the advent of new radio facilities, will enable an exploitation of this new diagnostic tool and making more use of radio observations to constrain the physics of GRBs.
\vspace{-0.5cm}
\section*{acknowledgments}
We thank Hendrik van Eerten, Lara Nava, Rodolfo Barniol Duran, Chryssa Kouveliotou and Sylvain Guiriec for helpful comments and suggestions.


\vspace{-0.5cm}
\begin{thebibliography}{}
	\makeatletter
	\relax
	\def\mn@urlcharsother{\let\do\@makeother \do\$\do\&\do\#\do\^\do\_\do\%\do\~}
	\def\mn@doi{\begingroup\mn@urlcharsother \@ifnextchar [ {\mn@doi@}
		{\mn@doi@[]}}
	\def\mn@doi@[#1]#2{\def\@tempa{#1}\ifx\@tempa\@empty \href
		{http://dx.doi.org/#2} {doi:#2}\else \href {http://dx.doi.org/#2} {#1}\fi
		\endgroup}
	\def\mn@eprint#1#2{\mn@eprint@#1:#2::\@nil}
	\def\mn@eprint@arXiv#1{\href {http://arxiv.org/abs/#1} {{\tt arXiv:#1}}}
	\def\mn@eprint@dblp#1{\href {http://dblp.uni-trier.de/rec/bibtex/#1.xml}
		{dblp:#1}}
	\def\mn@eprint@#1:#2:#3:#4\@nil{\def\@tempa {#1}\def\@tempb {#2}\def\@tempc
		{#3}\ifx \@tempc \@empty \let \@tempc \@tempb \let \@tempb \@tempa \fi \ifx
		\@tempb \@empty \def\@tempb {arXiv}\fi \@ifundefined
		{mn@eprint@\@tempb}{\@tempb:\@tempc}{\expandafter \expandafter \csname
			mn@eprint@\@tempb\endcsname \expandafter{\@tempc}}}
	
	\bibitem[\protect\citeauthoryear{{Abdo} et~al.,}{{Abdo}
		et~al.}{2009a}]{Abdo2009a}
	{Abdo} A.~A.,  et~al., 2009a, \mn@doi [Science] {10.1126/science.1169101},
	\href {http://adsabs.harvard.edu/abs/2009Sci...323.1688A} {323, 1688}
	
	\bibitem[\protect\citeauthoryear{{Abdo} et~al.,}{{Abdo}
		et~al.}{2009b}]{Abdo2009b}
	{Abdo} A.~A.,  et~al., 2009b, \mn@doi [\apjl] {10.1088/0004-637X/706/1/L138},
	\href {http://adsabs.harvard.edu/abs/2009ApJ...706L.138A} {706, L138}
	
	\bibitem[\protect\citeauthoryear{{Anderson} et~al.,}{{Anderson}
		et~al.}{2014}]{Anderson2014}
	{Anderson} G.~E.,  et~al., 2014, \mn@doi [\mnras] {10.1093/mnras/stu478}, \href
	{http://adsabs.harvard.edu/abs/2014MNRAS.440.2059A} {440, 2059}
	
	\bibitem[\protect\citeauthoryear{{Barniol Duran}}{{Barniol
			Duran}}{2014}]{BD2014}
	{Barniol Duran} R.,  2014, \mn@doi [\mnras] {10.1093/mnras/stu1070}, \href
	{http://adsabs.harvard.edu/abs/2014MNRAS.442.3147B} {442, 3147}
	
	\bibitem[\protect\citeauthoryear{{Beniamini} \& {Mochkovitch}}{{Beniamini} \&
		{Mochkovitch}}{2017}]{BM2017}
	{Beniamini} P.,  {Mochkovitch} R.,  2017, \mn@doi [\aap]
	{10.1051/0004-6361/201730523}, \href
	{http://adsabs.harvard.edu/abs/2017A%26A...605A..60B} {605, A60}
		
		\bibitem[\protect\citeauthoryear{{Beniamini}, {Nava}, {Duran}  \&
			{Piran}}{{Beniamini} et~al.}{2015}]{Beniamini2015}
		{Beniamini} P.,  {Nava} L.,  {Duran} R.~B.,   {Piran} T.,  2015, \mn@doi
		[\mnras] {10.1093/mnras/stv2033}, \href
		{http://adsabs.harvard.edu/abs/2015MNRAS.454.1073B} {454, 1073}
		
		\bibitem[\protect\citeauthoryear{{Beniamini}, {Nava}  \& {Piran}}{{Beniamini}
			et~al.}{2016}]{Beniamini2016}
		{Beniamini} P.,  {Nava} L.,   {Piran} T.,  2016, \mn@doi [\mnras]
		{10.1093/mnras/stw1331}, \href
		{http://adsabs.harvard.edu/abs/2016MNRAS.461...51B} {461, 51}
		
		\bibitem[\protect\citeauthoryear{{Burlon}, {Ghirlanda}, {van der Horst},
			{Murphy}, {Wijers}, {Gaensler}, {Ghisellini}  \& {Prandoni}}{{Burlon}
			et~al.}{2015}]{Burlon2015}
		{Burlon} D.,  {Ghirlanda} G.,  {van der Horst} A.,  {Murphy} T.,  {Wijers}
		R.~A.~M.~J.,  {Gaensler} B.,  {Ghisellini} G.,   {Prandoni} I.,  2015,
		Advancing Astrophysics with the Square Kilometre Array (AASKA14), \href
		{http://adsabs.harvard.edu/abs/2015aska.confE..52B} {p.~52}
		
		\bibitem[\protect\citeauthoryear{{Cenko} et~al.,}{{Cenko}
			et~al.}{2010}]{Cenko2010}
		{Cenko} S.~B.,  et~al., 2010, \mn@doi [\apj] {10.1088/0004-637X/711/2/641},
		\href {http://adsabs.harvard.edu/abs/2010ApJ...711..641C} {711, 641}
		
		\bibitem[\protect\citeauthoryear{{Cenko} et~al.,}{{Cenko}
			et~al.}{2011}]{Cenko2011}
		{Cenko} S.~B.,  et~al., 2011, \mn@doi [\apj] {10.1088/0004-637X/732/1/29},
		\href {http://adsabs.harvard.edu/abs/2011ApJ...732...29C} {732, 29}
		
		\bibitem[\protect\citeauthoryear{{Chandra} \& {Frail}}{{Chandra} \&
			{Frail}}{2012}]{CF2012}
		{Chandra} P.,  {Frail} D.~A.,  2012, \mn@doi [\apj]
		{10.1088/0004-637X/746/2/156}, \href
		{http://adsabs.harvard.edu/abs/2012ApJ...746..156C} {746, 156}
		
		\bibitem[\protect\citeauthoryear{{Chevalier} \& {Li}}{{Chevalier} \&
			{Li}}{1999}]{Chevalier1999}
		{Chevalier} R.~A.,  {Li} Z.-Y.,  1999, \mn@doi [\apjl] {10.1086/312147}, \href
		{http://adsabs.harvard.edu/abs/1999ApJ...520L..29C} {520, L29}
		
		\bibitem[\protect\citeauthoryear{{Chevalier} \& {Li}}{{Chevalier} \&
			{Li}}{2000}]{Chevalier2000}
		{Chevalier} R.~A.,  {Li} Z.-Y.,  2000, \mn@doi [\apj] {10.1086/308914}, \href
		{http://adsabs.harvard.edu/abs/2000ApJ...536..195C} {536, 195}
		
		\bibitem[\protect\citeauthoryear{{Costa} et~al.,}{{Costa}
			et~al.}{1997}]{Costa1997}
		{Costa} E.,  et~al., 1997, \mn@doi [\nat] {10.1038/42885}, \href
		{http://adsabs.harvard.edu/abs/1997Natur.387..783C} {387, 783}
		
		\bibitem[\protect\citeauthoryear{{De Colle}, {Ramirez-Ruiz}, {Granot}  \&
			{Lopez-Camara}}{{De Colle} et~al.}{2012}]{DeColle2012}
		{De Colle} F.,  {Ramirez-Ruiz} E.,  {Granot} J.,   {Lopez-Camara} D.,  2012,
		\mn@doi [\apj] {10.1088/0004-637X/751/1/57}, \href
		{http://adsabs.harvard.edu/abs/2012ApJ...751...57D} {751, 57}
		
		\bibitem[\protect\citeauthoryear{{Eichler} \& {Granot}}{{Eichler} \&
			{Granot}}{2006}]{EichlerGranot2006}
		{Eichler} D.,  {Granot} J.,  2006, \mn@doi [\apjl] {10.1086/503667}, \href
		{http://adsabs.harvard.edu/abs/2006ApJ...641L...5E} {641, L5}
		
		\bibitem[\protect\citeauthoryear{{Eichler} \& {Waxman}}{{Eichler} \&
			{Waxman}}{2005}]{Eichler2005}
		{Eichler} D.,  {Waxman} E.,  2005, \mn@doi [\apj] {10.1086/430596}, \href
		{http://adsabs.harvard.edu/abs/2005ApJ...627..861E} {627, 861}
		
		\bibitem[\protect\citeauthoryear{{Fan} \& {Piran}}{{Fan} \&
			{Piran}}{2006}]{Fan2006}
		{Fan} Y.,  {Piran} T.,  2006, \mn@doi [\mnras]
		{10.1111/j.1365-2966.2006.10280.x}, \href
		{http://adsabs.harvard.edu/abs/2006MNRAS.369..197F} {369, 197}
		
		\bibitem[\protect\citeauthoryear{{Frail}, {Kulkarni}, {Nicastro}, {Feroci}  \&
			{Taylor}}{{Frail} et~al.}{1997}]{Frail1997}
		{Frail} D.~A.,  {Kulkarni} S.~R.,  {Nicastro} L.,  {Feroci} M.,   {Taylor}
		G.~B.,  1997, \mn@doi [\nat] {10.1038/38451}, \href
		{http://adsabs.harvard.edu/abs/1997Natur.389..261F} {389, 261}
		
		\bibitem[\protect\citeauthoryear{{Frail}, {Waxman}  \& {Kulkarni}}{{Frail}
			et~al.}{2000}]{Frail2000}
		{Frail} D.~A.,  {Waxman} E.,   {Kulkarni} S.~R.,  2000, \mn@doi [\apj]
		{10.1086/309024}, \href {http://adsabs.harvard.edu/abs/2000ApJ...537..191F}
		{537, 191}
		
		\bibitem[\protect\citeauthoryear{{Frail} et~al.,}{{Frail}
			et~al.}{2001}]{Frail2001}
		{Frail} D.~A.,  et~al., 2001, \mn@doi [\apjl] {10.1086/338119}, \href
		{http://adsabs.harvard.edu/abs/2001ApJ...562L..55F} {562, L55}
		
		\bibitem[\protect\citeauthoryear{{Galama} et~al.,}{{Galama}
			et~al.}{1998}]{Galama1998}
		{Galama} T.~J.,  et~al., 1998, \mn@doi [\apjl] {10.1086/311424}, \href
		{http://adsabs.harvard.edu/abs/1998ApJ...500L.101G} {500, L101}
		
		\bibitem[\protect\citeauthoryear{{Gao} \& {Dai}}{{Gao} \&
			{Dai}}{2010}]{Gao2010}
		{Gao} Y.,  {Dai} Z.-G.,  2010, \mn@doi [Research in Astronomy and Astrophysics]
		{10.1088/1674-4527/10/2/005}, \href
		{http://adsabs.harvard.edu/abs/2010RAA....10..142G} {10, 142}
		
		\bibitem[\protect\citeauthoryear{{Gehrels} et~al.,}{{Gehrels}
			et~al.}{2004}]{Gehrels2004}
		{Gehrels} N.,  et~al., 2004, \mn@doi [\apj] {10.1086/422091}, \href
		{http://adsabs.harvard.edu/abs/2004ApJ...611.1005G} {611, 1005}
		
		\bibitem[\protect\citeauthoryear{{Genet}, {Daigne}  \& {Mochkovitch}}{{Genet}
			et~al.}{2007}]{Genet2007}
		{Genet} F.,  {Daigne} F.,   {Mochkovitch} R.,  2007, \mn@doi [\mnras]
		{10.1111/j.1365-2966.2007.12243.x}, \href
		{http://adsabs.harvard.edu/abs/2007MNRAS.381..732G} {381, 732}
		
		\bibitem[\protect\citeauthoryear{{Ghirlanda} et~al.,}{{Ghirlanda}
			et~al.}{2013}]{Ghirlanda2013}
		{Ghirlanda} G.,  et~al., 2013, \mn@doi [\mnras] {10.1093/mnras/stt1466}, \href
		{http://adsabs.harvard.edu/abs/2013MNRAS.435.2543G} {435, 2543}
		
		\bibitem[\protect\citeauthoryear{{Goldstein}, {Connaughton}, {Briggs}  \&
			{Burns}}{{Goldstein} et~al.}{2016}]{Goldstein2016}
		{Goldstein} A.,  {Connaughton} V.,  {Briggs} M.~S.,   {Burns} E.,  2016,
		\mn@doi [\apj] {10.3847/0004-637X/818/1/18}, \href
		{http://adsabs.harvard.edu/abs/2016ApJ...818...18G} {818, 18}
		
		\bibitem[\protect\citeauthoryear{{Goodman}}{{Goodman}}{1997}]{Goodman1997}
		{Goodman} J.,  1997, \mn@doi [\na] {10.1016/S1384-1076(97)00031-6}, \href
		{http://adsabs.harvard.edu/abs/1997NewA....2..449G} {2, 449}
		
		\bibitem[\protect\citeauthoryear{{Granot} \& {Sari}}{{Granot} \&
			{Sari}}{2002}]{GS2002}
		{Granot} J.,  {Sari} R.,  2002, \mn@doi [\apj] {10.1086/338966}, \href
		{http://adsabs.harvard.edu/abs/2002ApJ...568..820G} {568, 820}
		
		\bibitem[\protect\citeauthoryear{{Granot} \& {van der Horst}}{{Granot} \& {van
				der Horst}}{2014}]{GvdH2014}
		{Granot} J.,  {van der Horst} A.~J.,  2014, \mn@doi [\pasa]
		{10.1017/pasa.2013.44}, \href
		{http://adsabs.harvard.edu/abs/2014PASA...31....8G} {31, e008}
		
		\bibitem[\protect\citeauthoryear{{Kulkarni} et~al.,}{{Kulkarni}
			et~al.}{1998}]{Kulkarni1998}
		{Kulkarni} S.~R.,  et~al., 1998, \mn@doi [\nat] {10.1038/27139}, \href
		{http://adsabs.harvard.edu/abs/1998Natur.395..663K} {395, 663}
		
		\bibitem[\protect\citeauthoryear{{Laskar} et~al.,}{{Laskar}
			et~al.}{2013}]{Laskar2013}
		{Laskar} T.,  et~al., 2013, \mn@doi [\apj] {10.1088/0004-637X/776/2/119}, \href
		{http://adsabs.harvard.edu/abs/2013ApJ...776..119L} {776, 119}
		
		\bibitem[\protect\citeauthoryear{{Lemoine}, {Li}  \& {Wang}}{{Lemoine}
			et~al.}{2013}]{Lemoine2013}
		{Lemoine} M.,  {Li} Z.,   {Wang} X.-Y.,  2013, \mn@doi [\mnras]
		{10.1093/mnras/stt1494}, \href
		{http://adsabs.harvard.edu/abs/2013MNRAS.435.3009L} {435, 3009}
		
		\bibitem[\protect\citeauthoryear{{Leventis}, {van der Horst}, {van Eerten}  \&
			{Wijers}}{{Leventis} et~al.}{2013}]{Leventis2013}
		{Leventis} K.,  {van der Horst} A.~J.,  {van Eerten} H.~J.,   {Wijers}
		R.~A.~M.~J.,  2013, \mn@doi [\mnras] {10.1093/mnras/stt226}, \href
		{http://adsabs.harvard.edu/abs/2013MNRAS.431.1026L} {431, 1026}
		
		\bibitem[\protect\citeauthoryear{{Nava} et~al.,}{{Nava}
			et~al.}{2014}]{Nava2014}
		{Nava} L.,  et~al., 2014, \mn@doi [\mnras] {10.1093/mnras/stu1451}, \href
		{http://adsabs.harvard.edu/abs/2014MNRAS.443.3578N} {443, 3578}
		
		\bibitem[\protect\citeauthoryear{{Nousek} et~al.,}{{Nousek}
			et~al.}{2006}]{Nousek2006}
		{Nousek} J.~A.,  et~al., 2006, \mn@doi [\apj] {10.1086/500724}, \href
		{http://adsabs.harvard.edu/abs/2006ApJ...642..389N} {642, 389}
		
		\bibitem[\protect\citeauthoryear{{O'Brien} et~al.,}{{O'Brien}
			et~al.}{2006}]{OBrien2006}
		{O'Brien} P.~T.,  et~al., 2006, \mn@doi [\apj] {10.1086/505457}, \href
		{http://adsabs.harvard.edu/abs/2006ApJ...647.1213O} {647, 1213}
		
		\bibitem[\protect\citeauthoryear{{Panaitescu} \& {Kumar}}{{Panaitescu} \&
			{Kumar}}{2001}]{PK2001}
		{Panaitescu} A.,  {Kumar} P.,  2001, \mn@doi [\apjl] {10.1086/324061}, \href
		{http://adsabs.harvard.edu/abs/2001ApJ...560L..49P} {560, L49}
		
		\bibitem[\protect\citeauthoryear{{Panaitescu} \& {Kumar}}{{Panaitescu} \&
			{Kumar}}{2002}]{PK2002}
		{Panaitescu} A.,  {Kumar} P.,  2002, \mn@doi [\apj] {10.1086/340094}, \href
		{http://adsabs.harvard.edu/abs/2002ApJ...571..779P} {571, 779}
		
		\bibitem[\protect\citeauthoryear{{Perley} et~al.,}{{Perley}
			et~al.}{2014}]{Perley2014}
		{Perley} D.~A.,  et~al., 2014, \mn@doi [\apj] {10.1088/0004-637X/781/1/37},
		\href {http://adsabs.harvard.edu/abs/2014ApJ...781...37P} {781, 37}
		
		\bibitem[\protect\citeauthoryear{{Rhoads}}{{Rhoads}}{1999}]{Rhoads1999}
		{Rhoads} J.~E.,  1999, \mn@doi [\apj] {10.1086/307907}, \href
		{http://adsabs.harvard.edu/abs/1999ApJ...525..737R} {525, 737}
		
		\bibitem[\protect\citeauthoryear{{Ryan}, {van Eerten}, {MacFadyen}  \&
			{Zhang}}{{Ryan} et~al.}{2015}]{Ryan2015}
		{Ryan} G.,  {van Eerten} H.,  {MacFadyen} A.,   {Zhang} B.-B.,  2015, \mn@doi
		[\apj] {10.1088/0004-637X/799/1/3}, \href
		{http://adsabs.harvard.edu/abs/2015ApJ...799....3R} {799, 3}
		
		\bibitem[\protect\citeauthoryear{{Santana}, {Barniol Duran}  \&
			{Kumar}}{{Santana} et~al.}{2014}]{Santana2014}
		{Santana} R.,  {Barniol Duran} R.,   {Kumar} P.,  2014, \mn@doi [\apj]
		{10.1088/0004-637X/785/1/29}, \href
		{http://adsabs.harvard.edu/abs/2014ApJ...785...29S} {785, 29}
		
		\bibitem[\protect\citeauthoryear{{Sari} \& {Esin}}{{Sari} \&
			{Esin}}{2001}]{SariEsin2001}
		{Sari} R.,  {Esin} A.~A.,  2001, \mn@doi [\apj] {10.1086/319003}, \href
		{http://adsabs.harvard.edu/abs/2001ApJ...548..787S} {548, 787}
		
		\bibitem[\protect\citeauthoryear{{Sari}, {Piran}  \& {Narayan}}{{Sari}
			et~al.}{1998}]{Sari1998}
		{Sari} R.,  {Piran} T.,   {Narayan} R.,  1998, \mn@doi [\apjl]
		{10.1086/311269}, \href {http://adsabs.harvard.edu/abs/1998ApJ...497L..17S}
		{497, L17}
		
		\bibitem[\protect\citeauthoryear{{Sari}, {Piran}  \& {Halpern}}{{Sari}
			et~al.}{1999}]{Sari1999}
		{Sari} R.,  {Piran} T.,   {Halpern} J.~P.,  1999, \mn@doi [\apjl]
		{10.1086/312109}, \href {http://adsabs.harvard.edu/abs/1999ApJ...519L..17S}
		{519, L17}
		
		\bibitem[\protect\citeauthoryear{{Sironi} \& {Spitkovsky}}{{Sironi} \&
			{Spitkovsky}}{2011}]{Sironi2011}
		{Sironi} L.,  {Spitkovsky} A.,  2011, \mn@doi [\apj]
		{10.1088/0004-637X/726/2/75}, \href
		{http://adsabs.harvard.edu/abs/2011ApJ...726...75S} {726, 75}
		
		\bibitem[\protect\citeauthoryear{{Soderberg} et~al.,}{{Soderberg}
			et~al.}{2004}]{Soderberg2004}
		{Soderberg} A.~M.,  et~al., 2004, \mn@doi [\nat] {10.1038/nature02757}, \href
		{http://adsabs.harvard.edu/abs/2004Natur.430..648S} {430, 648}
		
		\bibitem[\protect\citeauthoryear{{Staley} et~al.,}{{Staley}
			et~al.}{2013}]{Stewart2013}
		{Staley} T.~D.,  et~al., 2013, \mn@doi [\mnras] {10.1093/mnras/sts259}, \href
		{http://adsabs.harvard.edu/abs/2013MNRAS.428.3114S} {428, 3114}
		
		\bibitem[\protect\citeauthoryear{{Uhm} \& {Beloborodov}}{{Uhm} \&
			{Beloborodov}}{2007}]{Uhm2007}
		{Uhm} Z.~L.,  {Beloborodov} A.~M.,  2007, \mn@doi [\apjl] {10.1086/519837},
		\href {http://adsabs.harvard.edu/abs/2007ApJ...665L..93U} {665, L93}
		
		\bibitem[\protect\citeauthoryear{{Wang}, {Liu}  \& {Lemoine}}{{Wang}
			et~al.}{2013}]{Wang2013}
		{Wang} X.-Y.,  {Liu} R.-Y.,   {Lemoine} M.,  2013, \mn@doi [\apjl]
		{10.1088/2041-8205/771/2/L33}, \href
		{http://adsabs.harvard.edu/abs/2013ApJ...771L..33W} {771, L33}
		
		\bibitem[\protect\citeauthoryear{{Warren}, {Ellison}, {Bykov}  \&
			{Lee}}{{Warren} et~al.}{2015}]{Warren2015}
		{Warren} D.~C.,  {Ellison} D.~C.,  {Bykov} A.~M.,   {Lee} S.-H.,  2015, \mn@doi
		[\mnras] {10.1093/mnras/stv1304}, \href
		{http://adsabs.harvard.edu/abs/2015MNRAS.452..431W} {452, 431}
		
		\bibitem[\protect\citeauthoryear{{Wijers} \& {Galama}}{{Wijers} \&
			{Galama}}{1999}]{WijersGalama1999}
		{Wijers} R.~A.~M.~J.,  {Galama} T.~J.,  1999, \mn@doi [\apj] {10.1086/307705},
		\href {http://adsabs.harvard.edu/abs/1999ApJ...523..177W} {523, 177}
		
		\bibitem[\protect\citeauthoryear{{Yost}, {Harrison}, {Sari}  \& {Frail}}{{Yost}
			et~al.}{2003}]{Yost2003}
		{Yost} S.~A.,  {Harrison} F.~A.,  {Sari} R.,   {Frail} D.~A.,  2003, \mn@doi
		[\apj] {10.1086/378288}, \href
		{http://adsabs.harvard.edu/abs/2003ApJ...597..459Y} {597, 459}
		
		\bibitem[\protect\citeauthoryear{{Zauderer} et~al.,}{{Zauderer}
			et~al.}{2013}]{Zauderer2013}
		{Zauderer} B.~A.,  et~al., 2013, \mn@doi [\apj] {10.1088/0004-637X/767/2/161},
		\href {http://adsabs.harvard.edu/abs/2013ApJ...767..161Z} {767, 161}
		
		\bibitem[\protect\citeauthoryear{{Zhang}, {van Eerten}, {Burrows}, {Ryan},
			{Evans}, {Racusin}, {Troja}  \& {MacFadyen}}{{Zhang}
			et~al.}{2015}]{Zhang2015}
		{Zhang} B.-B.,  {van Eerten} H.,  {Burrows} D.~N.,  {Ryan} G.~S.,  {Evans}
		P.~A.,  {Racusin} J.~L.,  {Troja} E.,   {MacFadyen} A.,  2015, \mn@doi [\apj]
		{10.1088/0004-637X/806/1/15}, \href
		{http://adsabs.harvard.edu/abs/2015ApJ...806...15Z} {806, 15}
		
		\bibitem[\protect\citeauthoryear{{Zou}, {Fan}  \& {Piran}}{{Zou}
			et~al.}{2009}]{Zou2009}
		{Zou} Y.-C.,  {Fan} Y.-Z.,   {Piran} T.,  2009, \mn@doi [\mnras]
		{10.1111/j.1365-2966.2009.14779.x}, \href
		{http://adsabs.harvard.edu/abs/2009MNRAS.396.1163Z} {396, 1163}
		
		\bibitem[\protect\citeauthoryear{{van Eerten}}{{van
				Eerten}}{2015}]{VanEerten2015}
		{van Eerten} H.~J.,  2015, \mn@doi [Journal of High Energy Astrophysics]
		{10.1016/j.jheap.2015.04.004}, \href
		{http://adsabs.harvard.edu/abs/2015JHEAp...7...23V} {7, 23}
		
		\bibitem[\protect\citeauthoryear{{van Eerten} \& {MacFadyen}}{{van Eerten} \&
			{MacFadyen}}{2013}]{VanEerten2013}
		{van Eerten} H.,  {MacFadyen} A.,  2013, \mn@doi [\apj]
		{10.1088/0004-637X/767/2/141}, \href
		{http://adsabs.harvard.edu/abs/2013ApJ...767..141V} {767, 141}
		
		\bibitem[\protect\citeauthoryear{{van Eerten}, {Zhang}  \& {MacFadyen}}{{van
				Eerten} et~al.}{2010}]{VanEerten2010}
		{van Eerten} H.,  {Zhang} W.,   {MacFadyen} A.,  2010, \mn@doi [\apj]
		{10.1088/0004-637X/722/1/235}, \href
		{http://adsabs.harvard.edu/abs/2010ApJ...722..235V} {722, 235}
		
		\bibitem[\protect\citeauthoryear{{van Eerten}, {Meliani}, {Wijers}  \&
			{Keppens}}{{van Eerten} et~al.}{2011}]{VanEerten2011}
		{van Eerten} H.~J.,  {Meliani} Z.,  {Wijers} R.~A.~M.~J.,   {Keppens} R.,
		2011, \mn@doi [\mnras] {10.1111/j.1365-2966.2010.17582.x}, \href
		{http://adsabs.harvard.edu/abs/2011MNRAS.410.2016V} {410, 2016}
		
		\bibitem[\protect\citeauthoryear{{van Eerten}, {van der Horst}  \&
			{MacFadyen}}{{van Eerten} et~al.}{2012}]{VanEerten2012}
		{van Eerten} H.,  {van der Horst} A.,   {MacFadyen} A.,  2012, \mn@doi [\apj]
		{10.1088/0004-637X/749/1/44}, \href
		{http://adsabs.harvard.edu/abs/2012ApJ...749...44V} {749, 44}
		
		\bibitem[\protect\citeauthoryear{{van Paradijs} et~al.,}{{van Paradijs}
			et~al.}{1997}]{VanParadijs1997}
		{van Paradijs} J.,  et~al., 1997, \mn@doi [\nat] {10.1038/386686a0}, \href
		{http://adsabs.harvard.edu/abs/1997Natur.386..686V} {386, 686}
		
		\bibitem[\protect\citeauthoryear{{van der Horst} et~al.,}{{van der Horst}
			et~al.}{2008}]{VanDerHorst2008}
		{van der Horst} A.~J.,  et~al., 2008, \mn@doi [\aap]
		{10.1051/0004-6361:20078051}, \href
		{http://adsabs.harvard.edu/abs/2008A%26A...480...35V} {480, 35}
			
			\bibitem[\protect\citeauthoryear{{van der Horst} et~al.,}{{van der Horst}
				et~al.}{2014}]{VanDerHorst2014}
			{van der Horst} A.~J.,  et~al., 2014, \mn@doi [\mnras] {10.1093/mnras/stu1664},
			\href {http://adsabs.harvard.edu/abs/2014MNRAS.444.3151V} {444, 3151}
			
			\makeatother
		\end{thebibliography}
\end{document}